\documentclass[12pt]{spieman}  
\usepackage{amsmath,amsfonts,amssymb}
\usepackage{graphicx}
\usepackage{setspace}
\usepackage{tocloft}
\usepackage{lineno}

\title{On the Effects of Pointing Jitter, Actuators Drift, Telescope Rolls
and Broadband Detectors in Dark Hole Maintenance and Electric Field
Order Reduction}

\author[a,*]{Leonid Pogorelyuk}
\author[b]{Laurent Pueyo}
\author[c,a]{N. Jeremy Kasdin}
\affil[a]{Princeton University, Department of Mechanical and Aerospace Engineering, 41 Olden St, Princeton, NJ, US, 08544}
\affil[b]{Space Telescope Science Institue, 3800 San Martin Drive Baltimore, MD, US, 21218}
\affil[c]{University of San Francisco, College of Arts and Sciences, 2130 Fulton St, San Francisco, CA, US, 94117}

\cftpagenumbersoff{figure}
\cftpagenumbersoff{table} 
\begin{document} 
\maketitle

\begin{abstract}
Space coronagraphs are projected to detect exoplantes that are at
least $10^{10}$ times dimmer than their host stars.
Yet, the actual detection threshold depends on the instrument's wavefront stability and varies by an order of magnitude with the choice of observation strategy and post-processing method.
In this paper the authors consider the performance of the previously introduced observation strategy (dark hole maintenance) and post-processing algorithm (electric field order reduction) in the presence of various realistic effects.
In particular, it will be shown that under some common
assumptions, the telescope's averaged pointing jitter translates into an additional light source
incoherent with the residual light from the star (speckles), and that
jitter ``modes'' can be identified in post-processing and distinguished
from a planet signal. We also show that the decrease in contrast due
to drift of voltages in deformable mirror actuators can be mitigated
by recursive estimation of the electric field in the high-contrast
region of the image (dark hole) using Electric Field Conjugation (EFC). Moreover, this can be done even
when the measured intensity is broadband, as long as it is well approximated
by an incoherent sum of monochromatic intensities. Finally, we assess
the performance of closed-loop vs.\ open-loop observation scenarios
through a numerical simulation of the Wide-Field Infra-Red Survey
Telescope (WFIRST). In particular, we compare the post-processing
factors of Angular Differential Imaging (ADI) with and  without Electric Field
Order Reduction (EFOR), which we extended to account for possible
telescope rolls and the presence of pointing jitter. For all observation parameters considered in this paper, close-loop dark hole maintenance resulted in significantly higher post-processing accuracy. 
\end{abstract}


{\noindent \footnotesize\textbf{*}Leonid Pogorelyuk,  \linkable{leonidp@princeton.edu} }

\begin{spacing}{2}   

\section{Introduction}

\subsection{Context}

Future space coronagraphs are projected to detect tens of exo-earths
that are $10^{10}$ times dimmer than their host stars.\cite{stark2019exoearth}
Contrasts of better than $4\cdot10^{-10}$ have already been demonstrated in the lab\cite{seo2019testbed}
in preparation for the WFIRST mission\cite{douglas2018wfirst,demers2018review}.
Such high contrasts are unlikely to persist on their own throughout lengthy
observations (tens of hours required to achieve a reasonable signal-to-noise ratio (S/N) to detect
a Jupiter-like planet\cite{nemati2017sensitivity}), due to thermal
and structural instabilities\cite{shaklan2011stability}. As an example,
the wavefront error budget for the quadrafoil error (4th Zernike mode)
is on the order of $0.1\:\mathrm{nm}$ for WFIRST\cite{zhou2018high},
and would be even lower for future missions\cite{garreth2017performance}
such as the Habitable Exoplanet Imaging Mission (HabEx)\cite{mennesson2016habitable}
and Large UV/Optical/IR Surveyor (LUVOIR)\cite{bolcar2017large}.

A potentially major source of instabilities are telescope maneuvers.
The proposed WFIRST observation scenario will keep the Deformable Mirrors (DMs) fixed for
8 hour long exposures followed by a 2 hour long dark hole ``re-creation''
procedure while pointing at a reference star\cite{bailey2018lessons}.
The electric field of the speckles will be estimated and reduced via
pair-probing and Electic Field Conjugation (EFC)\cite{give2011pair,give2007electric}
based on narrowband intensity measurement (either using an Integral
Field Spectrometer (IFS) or sequentially, using multiple narrowband filters).
This results in a sawtooth temporal pattern for the contrast
(see Figs.\ ~15,16 in \cite{krist2018wfirst}) with ``spikes'' due
to increased pointing jitter after the telescope ``switches'' back from the reference star.

An alternative approach (proposed by the authors in Ref.~\citenum{pogorelyuk2019dark}) is to maintain the contrast in the dark hole throughout the observation in a closed-loop fashion. 
In section~\ref{sec:maintenance}, we extend this method to work with broadband intensity measurements, illustrate how it handles pointing jitter (residual errors from fast-steering mirrors\cite{shi2019wfirst}) and drift of the DM actuators\cite{prada2019high}. The results are
numerically compared to the WFIRST observation scenario.

The authors have also previously suggested that known DM perturbations introduced during a closed-loop observation phase can be exploited in post-processing.
In section~\ref{sec:EFOR}, we extend Electric Field
Order Reduction (EFOR\cite{pogorelyuk2019reduced}) to incorporate images
taken at different telescope orientations and subtracts the contribution of the jitter (by identifying a small number of fast-varying jitter modes). We then estimate the post-processing errors
associated with open- and closed-loop observations, EFOR and Angular
Differential Imaging (ADI \cite{lowrance1998coronagraphic,marois2006angular}).

The numerical results in sections~\ref{sec:maintenance} and~\ref{sec:EFOR} suggest that dark hole maintenance increases the average contrast regardless of the nature of the drift, presence or absence of jitter and measurement type (monochromatic or broadband). Consequently, the planet detection thresholds are lower for data obtained in closed-loop scenarios, although the best choice of post-processing method depends on the parameters of the observation.

\subsection{Notations}
To generalize the discussion to both monochromatic and broadband light,
we discretize the time depended electric field of the speckles in
the high-contrast region of the image (the dark hole), $E(t,\mathbf{x},\lambda)$,
based on location, $\mathbf{x}$, and wavelength, $\lambda$ (similarly to the formulation in Ref.~\citenum{pogorelyuk2019reduced}). In other
words, the electric field in the focal plane will be modelled as 
\begin{equation}
\mathbf{E}(t)=\begin{bmatrix}\mathrm{Re}\left\{ E(t,\mathbf{x}_{1},\lambda_{1})\right\} \\
\mathrm{Im}\left\{ E(t,\mathbf{x}_{1},\lambda_{1})\right\} \\
\mathrm{Re}\left\{ E(t,\mathbf{x}_{1},\lambda_{2})\right\} \\
\mathrm{Im}\left\{ E(t,\mathbf{x}_{1},\lambda_{2})\right\} \\
\vdots
\end{bmatrix}\in\mathbb{R}^{2nl}\label{eq:E_discrete}
\end{equation}
where $\mathbf{x}_{p},\:1\le p\le n$ are the centers of the $n$ pixels
in the dark hole and $\lambda_{q},\:1\le q\le l$ is some discretization
of the spectrum. The intensity of the speckles at the detectors is
therefore given by
\begin{equation}
\mathbf{I}(t)=B\cdot\left(\mathbf{E}(t)\circ\mathbf{E}(t)\right)\in\mathbb{R}^{nm},\label{eq:B_def}
\end{equation}
where $\circ$ stands for the Hadamard product, $m$ is the number
of channels in the detector and $B\in\mathbb{R}^{nm\times2nl}$ is
the linear operator for summing the real and imaginary parts of
all wavelengths in a channel (for broadband light, $l>1$, and for a single
channel detector, $m=1$, it is given by the Kronecker product of the
identity matrix and a row vector of ones, $B=I_{n\times n}\otimes\mathbf{1}_{2l}^{T}$).

\section{\label{sec:maintenance}Dark Hole Maintenance}
Coronagraphs use DMs to achieve a high \textit{initial} contrast, for example, by measuring the focal
plane intensity with different control settings (probes \cite{give2011pair}),
estimating the electric field and applying corrections\cite{give2007electric}.
Most such approaches\cite{jovanovic2018review} do not incorporate a noise model and
therefore become extremely inaccurate when pointing at a dim star
since the number of photons detected per observation frame is of order
1 (and so is the noise variance). In contrast, recursive estimation via the Extended Kalman Fitler
(EKF)\cite{riggs2014optimal}, incorporates all
prior measurement and gives the appropriate weight to the new noisy
observations. It allows observing a dim target star while \textit{maintaining}
contrast only slightly lower than in a ``perfect'' dark hole\cite{pogorelyuk2019dark}.
Below we formulate the EKF for the general spectrum discretization, eqs.~(\ref{eq:E_discrete})-(\ref{eq:B_def}),
and address the effects of pointing jitter and DM actuators drift.

\subsection{\label{sub:jitter}Appearance of Pointing Jitter}

The pointing jitter due to telescope reaction wheels is mostly compensated
for by fast-steering mirrors (FSMs), but the residual wavefront perturbations are non-negligible\cite{shi2019wfirst}.
Unlike fluctuations in structural modes, their timescale is much shorter than a single exposure; hence, they cannot be temporally resolved. Instead, one has to consider
their averaged contribution to incoherent sources and assume that it
varies slowly across multiple frames.

We begin by splitting the temporal variations of the electric field
of the speckles into their average and zero mean components during
each observation frame $k$,
\begin{equation}
\mathbf{E}(t)=\left\langle \mathbf{E}\right\rangle (k)+\delta\mathbf{E}(t),\:t\in\left[t_{k},t_{k+1}\right),\label{eq:E_split}
\end{equation}
where $\left\langle \cdot\right\rangle (k)$ denotes averaging over
$\left[t_{k},t_{k+1}\right)$ and $\left\langle \delta\mathbf{E}\right\rangle (k)=\mathbf{0}$.
We further denote the sensitivity of the field to small control actuations,
$\mathbf{u}$, as the Jacobian $G^{U}=\partial\mathbf{E}/\partial\mathbf{u}$,
and \emph{define} the ``open-loop electric field'' as
\begin{equation}
\mathbf{E}^{OL}(k)\equiv\left\langle \mathbf{E}\right\rangle (k)-G^{U}\mathbf{u}(k).\label{eq:E_OL_def}
\end{equation}
In a perfectly linear model, $\mathbf{E}^{OL}(k)$ would be the average
electric field (including FSM corrections) if the DMs were kept fixed ($\mathbf{u}(k)=0$) after
the dark hole has been created.

The average speckle intensity, $\left\langle \mathbf{I}\right\rangle (k)$,
can be shown to have two components,
\begin{alignat}{1}
\left\langle \mathbf{I}\right\rangle (k) & =\mathbf{I}^{S}(k)+\mathbf{I}^{J}(k),\label{eq:I_avg_def}\\
\mathbf{I}^{S}(k) & =B\cdot\left(\mathbf{E}^{OL}(k)+G^{U}\mathbf{u}(k)\right)^{\circ2},\\
\mathbf{I}^{J}(k) & =\left\langle B\cdot\left(\delta\mathbf{E}(t)\right)^{\circ2}\right\rangle (k).\label{eq:I_J_def}
\end{alignat}
Here $\mathbf{I}^{S}(k)$ captures (mostly) the variations of the
speckles due to ``slow'' thermal and structural instabilities, while
$\mathbf{I}^{J}(k)$ is mostly due to jitter (the time average of the cross term $\left(\mathbf{E}^{OL}(k)+G^{U}\mathbf{u}(k)\right)\circ\delta\mathbf{E}(t)$
  is zero). The total intensity,
$\mathbf{I}$, also includes sources incoherent with the star, $\mathbf{I}^{I}$,
and detector noise sources, $\mathbf{I}^{D}$ (e.g. dark current),
\begin{equation}
\mathbf{I}=\mathbf{I}^{S}+\mathbf{I}^{J}+\mathbf{I}^{I}+\mathbf{I}^{D}.\label{eq:I_def}
\end{equation}

It is now evident that the fast variations in the electric field
due to jitter resemble other incoherent sources in the sense that
they remain almost unaffected by slow variations in high-order wavefront
errors and DM commands. Differentiating between jitter and other incoherent
sources is done in post-processing (Sec.~\ref{sub:jitter_rolls})
while the algorithm below allows for a reduction in the intensity of the speckles,
$\mathbf{I}^{S}$---the only source affected by the DMs.

\subsection{Broadband Extended Kalman Filter}

\begin{figure}
\begin{center}
\begin{tabular}{c}
\includegraphics[width=14cm]{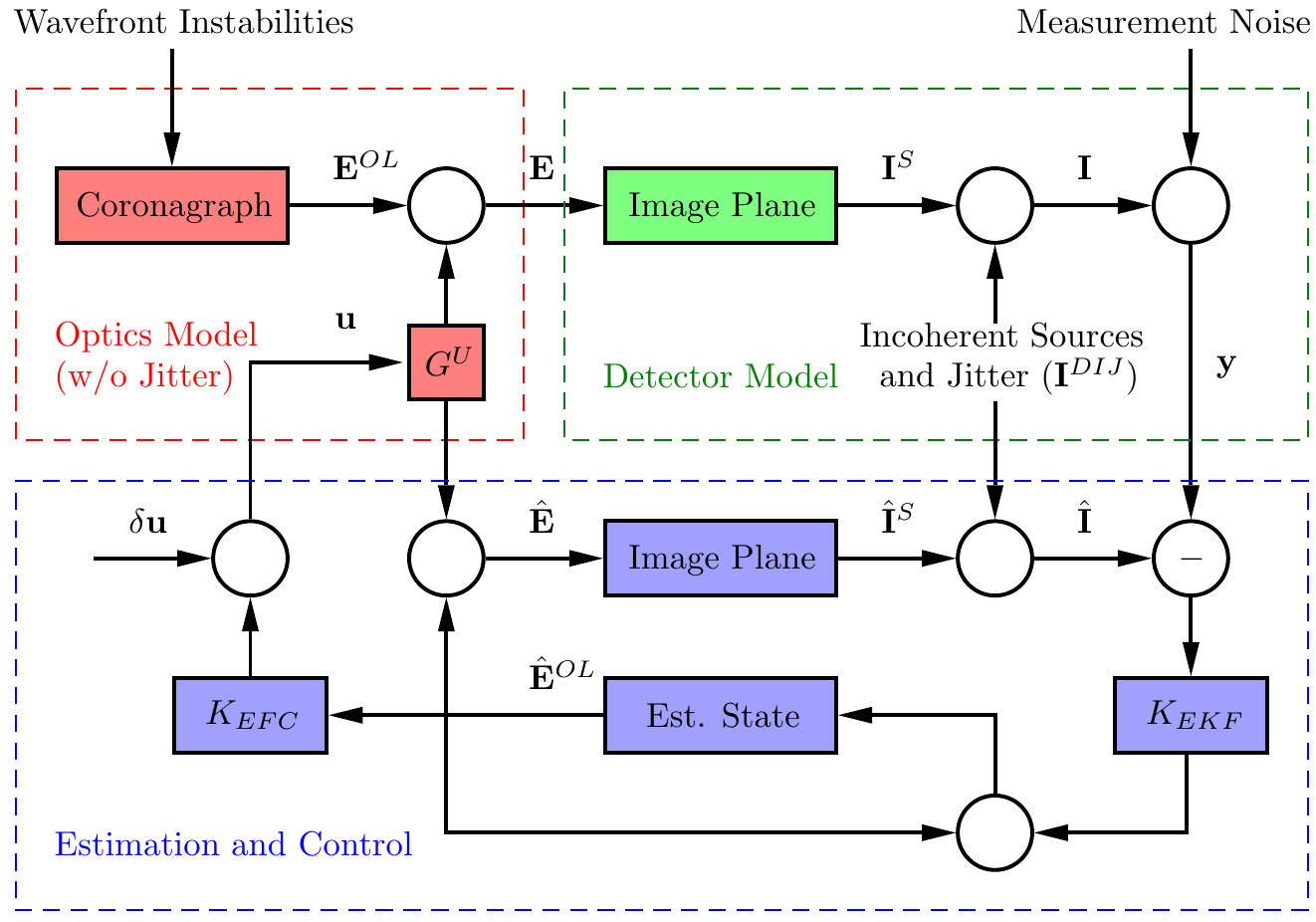}
\end{tabular}
\end{center}
\caption{\label{fig:diagram}A diagram of the \textit{linearized} model of the coronagraph with jitter contribution approximated by an incoherent light source. The top half of this figure describes how the actual instrument hardware is simulated in this paper while the bottom half illustrates the main functionalities of the Dark hole maintenance algorithm.}
\end{figure}

We are interested in efficiently computing a rough estimate of the
slowly varying open-loop electric field, $\hat{\mathbf{E}}^{OL}$,
in order to apply corrections to the DMs,
\begin{equation}
\mathbf{u}(k)=-K_{EFC}\hat{\mathbf{E}}^{OL}(k-1)+\delta\mathbf{u}(k),\label{eq:closed_loop}
\end{equation}
where $K_{EFC}$ is the EFC gain\cite{give2007electric}, and $\delta\mathbf{u}(k)$ is random DM dither (Fig.~\ref{fig:diagram}).
Note that the mean component of the jitter is already included in $\mathbf{E}^{OL}$ and we further assume that ``variance'' of the jitter, $\mathbf{I}^{J}(k)$, changes slowly between frames. We may therefore gather the terms in Eq.~(\ref{eq:I_def})
which are unaffected by controls
\begin{equation}
\mathbf{I}(k)=\mathbf{I}^{S}(k)+\mathbf{I}^{DIJ}(k),
\end{equation}
and apply the existing EKF algorithm for estimating both $\mathbf{E}^{OL}(k)$
and $\mathbf{I}^{DIJ}(k)$ without modifications (see Appendix A in Ref.~\citenum{pogorelyuk2019dark}).
However, since the variations in the residual pointing jitter are very small, the authors found $\mathbf{I}^{DIJ}\approx\mathrm{const}$
to be empirically justifiable as it reduces the dimension of the
EKF, making it more tractable without compromising its accuracy.

Another simplification, for numerical purposes, consists of ignoring
the correlation of electric field increments between pixels. We assume that they are normally distributed with zero mean and a block diagonal covariance matrix, $\Sigma$,
\begin{equation}
\mathbf{E}^{OL}(k)-\mathbf{E}^{OL}(k-1)\sim{\cal N}\left(\mathbf{0},\Sigma\right).\label{eq:Sigma_def}
\end{equation}
Ignoring the dependence between pixels (by discarding the off-diagonal terms in $\Sigma$) greatly
reduces the accuracy of the filter but allows propagating all the
estimates in ${\cal O}(n)$ operations per frame instead of the ${\cal O}(n^{3}$)
required for the full EKF (it is possible to exploit the low-order
nature of the electric field increments to recover some of the information
while keeping an ${\cal O}(n)$ complexity\cite{pogorelyuk2020efficient}).

The EKF is formulated in terms of the open-loop electric field estimates,
$\hat{\mathbf{E}}^{OL}(k)$. It is advanced based on the the number
of photons detected in the physical system, $\mathbf{y}(k)$, which presumably follows a multivariate
Poisson distribution parameterized by the intensity, $\mathbf{I}(k)$.
For \textit{estimation} purposes, it is approximated by a normal distribution
with mean and variance both equal to the intensity estimate, $\mathbf{\hat{\mathbf{I}}}$,
at step $k$,
\begin{alignat}{1}
\mathbf{y}(k) & \sim{\cal N}\left(\hat{\mathbf{I}}(k),\mathrm{diag}\left(\hat{\mathbf{I}}(k)\right)\right),\\
\hat{\mathbf{I}}(k) & =B\cdot\left(\hat{\mathbf{E}}^{OL}(k)+G^{U}\mathbf{u}(k)\right)^{\circ2}+\mathbf{I}^{DIJ},\label{eq:I_hat_def}
\end{alignat}
where $\mathrm{diag}(\mathbf{a})$ stands for a diagonal matrix with
the elements of $\mathbf{a}$ on its diagonal. 

Apart from the shot noise, which has an estimated covariance of $\mathrm{diag}\left(\hat{\mathbf{I}}(k)\right)$, the uncertainty in the state estimate, $\hat{\mathbf{E}}^{OL}(k)-\mathbf{E}^{OL}(k)$, also has a covariance, $P(k)$, which needs to be accounted for.
The two are combined in the EKF gain given by
\begin{equation}
K_{EKF}(k) = P(k)H^{T}(k)\left(H(k)P(k)H^{T}(k)+\mathrm{diag}\left(\hat{\mathbf{I}}(k)\right)\right)^{-1},
\end{equation}
where $H(k)$ denotes the linearized effect of the open-loop electric field on the measurements,
\begin{equation}
H(k)=\frac{\partial\hat{\mathbf{I}}(k)}{\partial\hat{\mathbf{E}}^{OL}(k)}=2B\cdot\mathrm{diag}\left(\hat{\mathbf{E}}^{OL}(k)+G^{U}\mathbf{u}(k)\right).\label{eq:H_def}
\end{equation}
Note that the last equation requires that both $\mathbf{u}(k)$ and $\hat{\mathbf{E}}^{OL}(k)-\mathbf{E}^{OL}(k)$ be small, so that the coronagraph is in the linear regime and the EKF gain is ``roughly in the correct direction''.

The above gain is used to advance the estimated state based on the discrepancy between the predicted measurement, $\hat{\mathbf{I}}(k)$, and the actual one, $\mathbf{y}(k)$, (the innovation),
\begin{equation}
\hat{\mathbf{E}}^{OL}(k+1)  =\hat{\mathbf{E}}^{OL}(k)+K_{EKF}(k)\left(\mathbf{y}(k)-\hat{\mathbf{I}}(k)\right).
\end{equation}
As a result, the gain acts to decrease the error covariance, $P$, which would otherwise be constantly increasing due to unknown drift increments (with covariance $\Sigma$). Both of these effects are accounted for when  advancing the covariance matrix approximation,
\begin{equation}
P(k+1) =\left(I-K_{EKF}(k)H(k)\right)P(k)+\Sigma.
\end{equation}

Since we ignore the cross-correlation between pixels, the matrices
$P,R,H$ and $\Sigma$ are block diagonal. Taking advantage of that,
the EKF can be advanced independently for each pixel giving an ${\cal O}(n)$
time and space complexity for all the pixels combined. We also note
that the spectral discretization in the above formulation is implicit
in the dimensions of $\mathbf{y}$, $\hat{\mathbf{E}}$ and the elements
of $B$. Hence, it applies to both to single- ($l>m=1$) and multi-
channel ($l\ge m>1$) detectors. 
Finally, we require that the magnitude (covariance) of the dither, $\delta\mathbf{u}$, is comparable to $\Sigma$ (although its exact distribution is not important if it is random).

\subsection{The Effect of DM Actuators Drift}

In a realistic scenario, the actual DM actuations, $\mathbf{u}_{true}(k)$,
might slightly differ from the deterministic commands, $\mathbf{u}(k)$\cite{prada2019high}.
If left unchecked, these DM surface discrepancies might result in an unwanted intensity buildup.

For lack of a better model, we assume that the time-evolution
of the difference between prescribed and actual commands, i.e., the DM drift, can be approximated by a random walk of a known magnitude, $\sigma_{a}$,
\begin{flalign}
\mathbf{u}_{true}(k) & =\mathbf{u}(k)+\mathbf{u}_{drift}(k),\label{eq:u_true_def}\\
\mathbf{u}_{drift}(k+1)-\mathbf{u}_{drift}(k) & \sim{\cal N}\left(\mathbf{0},\sigma_{a}^{2}I\right).\label{eq:u_drift_def}
\end{flalign}
Given the definition of the open-loop electric field, (\ref{eq:E_OL_def}),
we observe that the only implication of the DM drift is the increase
in the covariance of the electric field increments, and in particular
\begin{equation}
\mathrm{cov}\left\{\mathbf{E}^{OL}(k+1)-\mathbf{E}^{OL}(k)\right\}\ge\sigma_{a}^{2}G^{U}\left(G^{U}\right)^{T}.
\end{equation}

While the EKF in eqs.~(\ref{eq:I_hat_def})-(\ref{eq:H_def}) remain
unchanged (regardless of whether the actuators are statistically independent or not), the full drift covariance and its block diagonal approximation, $\Sigma$, have
to be increased to account for this additional source of uncertainty.
It's worth noting that the accuracy of the filter is not very sensitive
to the exact values of the elements of $\Sigma$.

\subsection{\label{sub:EKF_numerical}Numerical Results}

To assess the performance of the dark hole maintenance scheme\cite{pogorelyuk2019dark} in the presence of the above mentioned effects, the authors employed the FALCO\cite{riggs2018fast} model of the WFIRST Hybrid Lyot Coronagraph. The simulations consisted of series of 5 single wavelength images
equally spaced between $546\:\mathrm{nm}$ and $601\:\mathrm{nm}$.
The initial dark hole had a contrast of $4.2\cdot10^{-10}$ in a ring
between $3$ and $9$ $\lambda/D$ with an average flux of $1.2\:\frac{\mathrm{\mathrm{photon}}}{\mathrm{frame}}$
(per pixel and wavelength) from the star and $0.25\:\frac{\mathrm{\mathrm{photon}}}{\mathrm{frame}}$
from dark current. This slowly-varying speckle drift was simulated via
random walk of the first $21$ Zernike polynomials (denoted by their time averages within each frame),
\begin{equation}
\left\langle z_{p}^{p-2j}\right\rangle (k+1)-\left\langle z_{p}^{p-2j}\right\rangle (k)\sim{\cal N}\left(0,\left(\frac{\sigma_{d}}{(p+1)^{2}\cdot\lambda}\right)^{2}\Delta t\right),\;0\le j\le p
\end{equation}
where $p$ is the order of the polynomial, $j$ is its azimuthal degree,
$\Delta z_{p}^{p-2j}$ is its increment over one $\Delta t=100\:\mathrm{sec}$
frame, and $\sigma_{d}=0.2\:\frac{\mathrm{nm}}{\sqrt{\mathrm{hr}}}$.
The pointing jitter in the simulation can be described as fast
periodic perturbations of the the tip/tilt Zernikes, $z_{\mathrm{tip/tilt}}=z_1^{\pm1}$,
\begin{alignat}{1}
z_{\mathrm{tip}}(t)-\left\langle z_{\mathrm{tip}}\right\rangle (k+1) & =a_{\mathrm{tip}}(k)\sin\left(\frac{2\pi}{\Delta t}t\right),\:t\in\left[t_{k},t_{k+1}\right),\label{eq:tip_jitter}\\
z_{\mathrm{tilt}}(t)-\left\langle z_{\mathrm{tilt}}\right\rangle (k+1) & =a_{\mathrm{tilt}}(k)\sin\left(\frac{2\pi}{\Delta t}t+\phi_{k}\right),\:t\in\left[t_{k},t_{k+1}\right),\label{eq:tilt_jitter}
\end{alignat}
where $a_{\mathrm{tip/tilt}}(k)$ varied slowly throughout the simulation between
$0$ and $1.4\:\mathrm{nm}$ and $\phi_{k}$ between $0$ and $2\pi$
(the added intensity was computed by analytically averaging Eq.~(\ref{eq:I_J_def})). 

The closed loop control law in Eq.~(\ref{eq:closed_loop}) was applied with $\delta\mathbf{u}(k)\sim{\cal N}\left(\mathbf{0},\sigma_{u}^{2}I\right)$ and $\sigma_{u}=5-10\:\mathrm{mV}$ (this dither introduces phase diversity and keeps the EKF stable and its optimal magnitude depends on the drift rate).
While the above control command was ``passed'' to the
EKF, the actual DM command used for the simulation also included actuators
drift with $\sigma_{a}=5\:\frac{\mathrm{mV}}{\sqrt{\mathrm{hr}}}$
as described in eqs.~(\ref{eq:u_true_def})-(\ref{eq:u_drift_def}) (this value was chosen such that DM and wavefront drift effects are of the same order).

\begin{figure}
\begin{center}
\begin{tabular}{c}
\includegraphics[width=14cm]{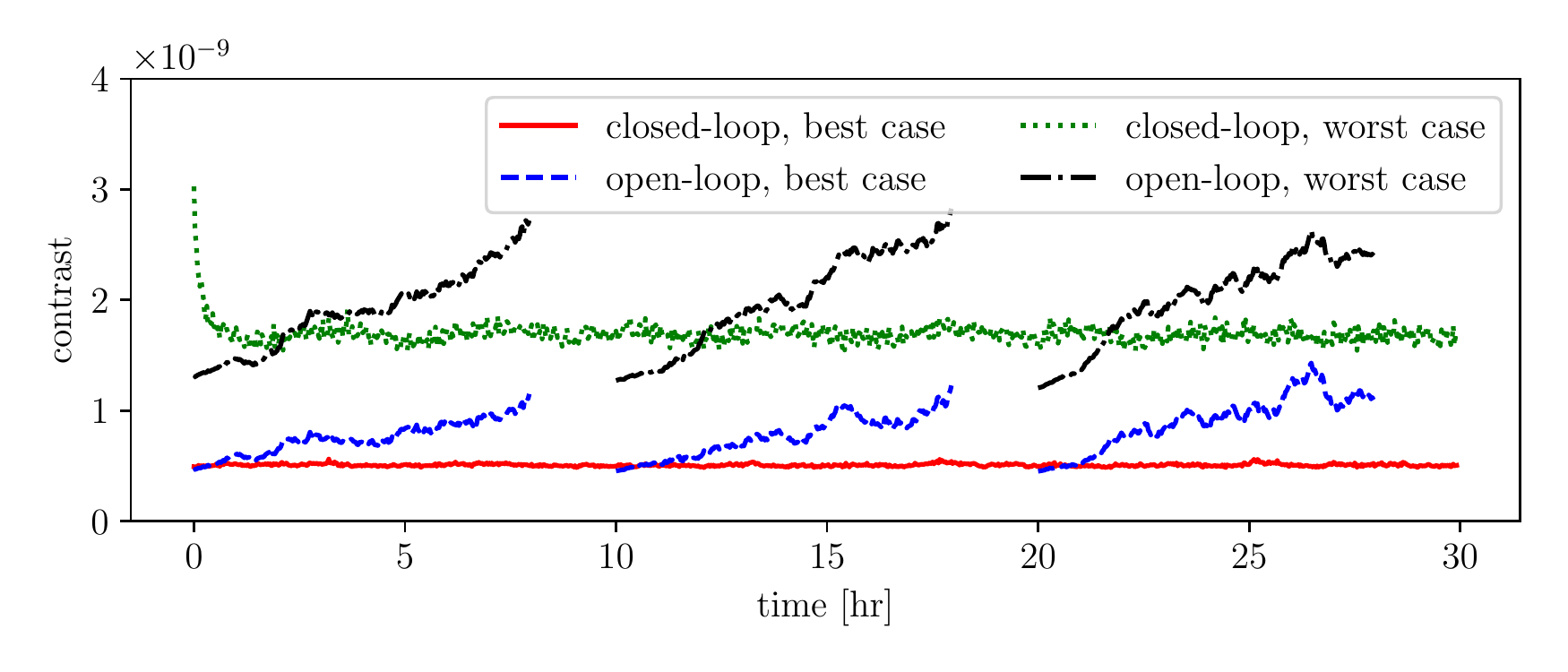}
\end{tabular}
\end{center}
\caption{\label{fig:EKF}Dark hole maintenance\cite{pogorelyuk2019dark} in the presence of realistic effects (dotted green line) and its comparison to an open loop
observation with periodic re-creation of the dark hole (dash-dotted black). In the best-case scenario (solid red and dashed
blue), only Zernike drift was present and the five narrowband
channels were available separately for the EKF. In the worst-case scenario (dotted
green and dash-dotted black), Zernikes, DM actuators drift and pointing jitter
were present and just one broadband channel was used by the EKF. As a result, the dither necessary to keep the EKF stable was higher and the contrast was worse.}
\end{figure}

Fig.~\ref{fig:EKF} compares the dark hole maintenance (closed-loop)
scheme with the proposed WFIRST-CGI open-loop observation scenario
which requires periodically observing a reference star to re-create
the dark hole\cite{bailey2018lessons} (hence the missing two-hour
segments in the open-loop lines). When only Zernikes drift was simulated
and all five channels were available for the EKF (best case),
a small dither magnitude, $\sigma_{u}=5\:\mathrm{mV}$, was sufficient
to maintain the contrast at almost its initial level. When DM drift
and jitter were present (worst case), the initial contrast was lower
and the open-loop contrast decreased significantly faster. Additionally,
the EKF had access to just the sum of all the channels, hence a larger
dither magnitude, $\sigma_{u}=10\:\mathrm{mV}$, was necessary to
ensure stability, it took more time to converge and the closed loop
contrast varied slightly with the jitter magnitude (dotted green line).

In all cases, the closed-loop approach maintained a significantly
higher contrast throughout most of the observation, compared to the
open-loop observation scenario. Besides utilizing close to 100\% of
the duty cycle, closing the loop also avoids telescope pointing maneuvers which might
excite structural modes and decrease the open-loop contrast further
(not simulated here).

Another potential benefit of closed-loop observations manifests itself in post-processing.
While DM dithering was introduced to stabilize the EKF, it also provides phase diversity
which helps to detect faint sources below the speckle floor. This is a non-linear and computationally expensive procedure, but it also allows accounting for the low-dimensionality of the speckles and telescope rolls as discussed in the next section.

\subsection{\label{sub:dark_hole_creation}Dark Hole Creation with Broadband
Measurements}

\begin{figure}
\begin{center}
\begin{tabular}{c}
\includegraphics[width=14cm]{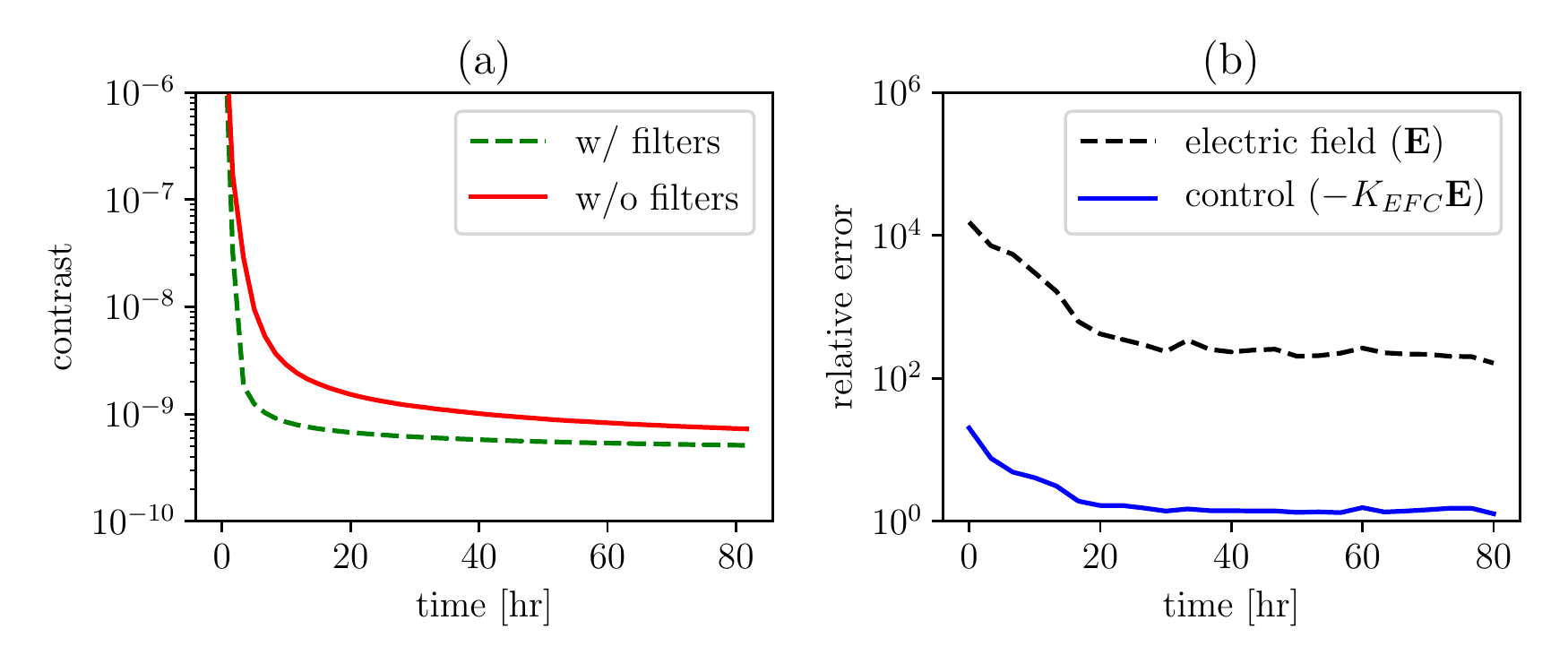}
\end{tabular}
\end{center}
\caption{\label{fig:EFC}(a) Contrast evolution during the creation of the
dark hole with pair-probing and EFC. When monochromatic intensities
where measured sequentially (using optical filters, dashed green line), the
contrast increased faster than when the all measurements were broadband
(without filters, solid red line). (b) The relative errors in the
electric field estimate (dashed black line) and the corresponding EFC
correction (solid blue line), both based on broadband measurements.
The ambiguities involved in estimating spectral information from broadband
measurements are ``cancelled out'' by the EFC gain for reasons discussed
in Appendix~\ref{sec:appendix}.}
\end{figure}

In the ``closed-loop worst-case'' simulation in figure \ref{fig:EKF},
the contrast was maintained with focal-plane broadband measurements
alone. This suggests that it could also be possible to \emph{create
}the dark hole without an IFS or optical filters.
Indeed, the pair-probing approach\cite{give2011pair} for estimating
the electric field can be modified slightly to include broadband measurements,
and the EFC control law for increasing the contrast is already ``broadband''.

The goal of pair-probing is to estimate the electric field in the
dark hole, $\mathbf{E}^{DH}$, by measuring the effects of $2s$ DM
probes $\left\{ \pm\delta\mathbf{u}_{j}\right\} _{j=1}^{s}$on the
intensity, $\mathbf{I}^{DH}$. It is assumed that the intensity measurements
have a high S/N and that the speckles and incoherent sources do not
drift during the ``probing''. With the generalized spectral discretization
in Eq.~(\ref{eq:E_discrete}), the (possibly broadband) intensity
is given by
\begin{equation}
\mathbf{I}^{DH}(\pm\delta\mathbf{u}_{j})=B\cdot\left(\mathbf{E}^{DH}\pm G^{U}\delta\mathbf{u}_{j}\right)^{\circ2}+\mathbf{I}^{DIJ},\label{eq:probing_I}
\end{equation}
where $\mathbf{I}^{DIJ}=\mathrm{const}$ is the sum of incoherent
sources. It follows that
\begin{equation}
\left(4B\cdot\mathrm{diag}(G^{U}\delta\mathbf{u}_{j})\right)\mathbf{E}^{DH}=\mathbf{I}^{DH}(+\delta\mathbf{u}_{j})-\mathbf{I}^{DH}(-\delta\mathbf{u}_{j}).
\end{equation}

Each pair of probes therefore gives $mn$ sparse linear equations
in $2nl=\mathrm{dim}\hat{\mathbf{E}}$ unknowns. Picking $s\ge2nl/m$
probes results in a least-squares problem for estimating $\mathbf{E}^{DH}$,
\begin{equation}
\begin{bmatrix}4B\cdot\mathrm{diag}(G^{U}\delta\mathbf{u}_{1})\\
\vdots\\
4B\cdot\mathrm{diag}(G^{U}\delta\mathbf{u}_{s})
\end{bmatrix}\hat{\mathbf{E}}^{DH}\approx\begin{bmatrix}\mathbf{I}^{DH}(+\delta\mathbf{u}_{1})-\mathbf{I}^{DH}(-\delta\mathbf{u}_{1})\\
\vdots\\
\mathbf{I}^{DH}(+\delta\mathbf{u}_{s})-\mathbf{I}^{DH}(-\delta\mathbf{u}_{s})
\end{bmatrix}.\label{eq:probing_estimate}
\end{equation}
The estimate, $\hat{\mathbf{E}}^{DH}$, is then used with the EFC control
law,
\begin{equation}
\Delta\mathbf{u}^{DH}=-K_{EFC}\hat{\mathbf{E}}^{DH},\label{eq:probing_control}
\end{equation}
where $\Delta\mathbf{u}^{DH}$ is the DM correction to be applied
before the next iteration of the dark hole creation sequence. Note
that the Jacobian, $G^{U}$, and the estimate, $\hat{\mathbf{E}}^{DH}$,
contain information about the spectrum while the measurements, $\mathbf{I}^{DH}$,
may now contain an arbitrary number of channels.

Figure~\ref{fig:EFC}(a) shows a dark hole creation sequence using
the same FALCO\cite{riggs2018fast} model of the WFIRST as the previous
subsection but slightly different observation parameters: no wavefront
drift or jitter and a 16 times brighter ``reference'' star. The
broadband variant of the algorithm, i.e., without optical filters
or $m=1$, was able to increase the contrast from $10^{-5}$ to $10^{-9}$.
Yet, it converged slower than the algorithm which measured each of
the five channels separately, i.e., $m=l=5$ (both variants are described
by eqs.~(\ref{eq:probing_estimate})-(\ref{eq:probing_control})
with a different $B$ matrix).

Since the dark holes were created using ``random'' probes and a fixed EFC gain, the convergence rates in Fig.~\ref{fig:EFC}(a) are qualitative. Yet, they demonstrate the feasibility of operating a purely broadband coronagraph while pointing out the advantage of having optical filters.

Arguably, it should be difficult to distinguish between speckles at
different wavelengths based on an incoherent sum of of their intensities.
To assess the accuracy of the electric field estimates in each wavelength,
Fig.~\ref{fig:EFC}(b) plots their relative error, $\left\Vert \mathbf{E}^{DH}-\hat{\mathbf{E}}^{DH}\right\Vert /\left\Vert \mathbf{E}^{DH}\right\Vert $.
As it turns out, the errors in $\hat{\mathbf{E}}^{DH}$ are many orders
of magnitude larger than the actual electric field, $\mathbf{E}^{DH}$,
although they are not propagated to $\Delta\mathbf{u}^{DH}$. In other words, the relative difference between
the ``perfect'' and estimated DM increments, $\left\Vert K_{EFC}\mathbf{E}^{DH}-K_{EFC}\hat{\mathbf{E}}^{DH}\right\Vert /\left\Vert K_{EFC}\mathbf{E}^{DH}\right\Vert $,
is small enough to allow creating the dark hole. This peculiar ``error
cancelling'' is analyzed in Appendix~\ref{sec:appendix}.

\section{\label{sec:EFOR}Electric Field Order Reduction}

The observation scenario proposed for WFIRST-CGI prescribes collecting
reference images every ten hours and $\pm13\:\mathrm{deg}$ telescope
rolls every two hours\cite{bailey2018lessons}. These maneuvers will
be utilized in post-processing via Angular Differential Imaging (ADI)\cite{lowrance1998coronagraphic,marois2006angular}
and Reference Differential Imaging (RDI)\cite{mawet2011dim}. The
collection of all the reference images could, in theory, produce a
library of ``most descriptive'' speckle patterns which could be
projected out of the observation\cite{soummer2012detection,amara2012pynpoint,ren2018non}.
However, the DM commands will change between the observations (either
due to drift or due to dark hole re-creation every ten hours), resulting
in temporally varying high-order wavefront aberrations. We expect this
to reduce the applicability of reference images between temporally
distant observations thus hindering the use of intensity-based order
reduction methods.

Another source of phase diversity is provided by the dithering of the DMs
during a closed-loop observation (see, Eq.~(\ref{eq:closed_loop})); this can help differentiate between speckles which are affected by
the DMs and other intensity sources\cite{pogorelyuk2019dark}. However,
as discussed in Sec.~\ref{sub:jitter}, the jitter is not affected
by dither and would be indistinguishable from incoherent sources
including planets. Nevertheless, the Electric Field Order Reduction (EFOR)
method proposed in  Ref.~\citenum{pogorelyuk2019reduced} makes additional
assumptions on the nature of the speckle drift that allow identifying
the jitter components in the images. In particular, it assumes that
the open-loop electric field lies in a low-dimensional subspace,
\begin{equation}
\mathbf{E}^{OL}=G^{V}\mathbf{v},
\end{equation}
where the columns of $G^{V}\in\mathbb{R}^{2nl\times r}$ form a basis
of $r$ ``significant'' electric field modes ($r$ depends on the
telescope configuration, e.g. segmented vs. monolithic, and is a tuning
parameter that can be optimized\cite{pogorelyuk2019maintaining}).
Below we describe EFOR and extend it to account for telescope rolls
and identify a small number of the modes in $G^{V}$ as ``jitter''
modes, thus allowing them to be subtracted from the incoherent signal estimate.
Our numerical simulations show that this low-order assumption breaks
down if the dominant source of wavefront drift are DM actuators (due
to their high spatial frequency), in which case a closed-loop observation
scenario with ADI gives the most accurate estimates.

\subsection{\label{sub:jitter_rolls}Pointing Jitter and Telescope Rolls}

Similarly to Eq.~(\ref{eq:E_split}), time variations of speckle
modes, $\mathbf{v}$, can be split into their average, $\left\langle \mathbf{v}\right\rangle $,
and zero mean, $\delta\mathbf{v}$, components,
\begin{equation}
\mathbf{v}(t)=\left\langle \mathbf{v}\right\rangle (k)+\delta\mathbf{v}(t),\:t\in\left[t_{k},t_{k+1}\right).
\end{equation}
Consequently, the jitter term defined in Eq.~(\ref{eq:I_J_def})
becomes
\begin{equation}
\mathbf{I}^{J}(k)=\left\langle B\cdot\left(G^{V}\delta\mathbf{v}(t)\right)^{\circ2}\right\rangle (k),
\end{equation}
and resides in a high-dimensional space spanned by ${r \choose 2}$
products of the $r$ columns of $G^{V}$. To reduce the number of
free parameters, we assume that the fast variations, $\delta\mathbf{v}$,
are negligible in all but $m$ of the modes (e.g. tip/tilt modes due
to pointing jitter, $m=2$ in eqs.~(\ref{eq:tip_jitter})-(\ref{eq:tilt_jitter})).
The contribution of the jitter, $\mathbf{I}^{J}(k)$, can then be
written as
\begin{equation}
\mathbf{I}^{J}(k)=B\cdot\underset{i=1}{\overset{m}{\sum}}\left(\underset{j=1}{\overset{i}{\sum}}w_{i,j}(k)\mathbf{g}_{j}^{V}\right)^{\circ2},\label{eq:jitter_modes}
\end{equation}
where $\mathbf{g}_{j}^{V}$ is the $j$th column of $G^{V}$ and $w_{i,j}(k)$
are some ${m \choose 2}$ coefficients (since $\mathbf{g}_{j}^{V}$,
$v_{i}(k)$ and $w_{i,j}(k)$ will be estimated simultaneously, the
choice of $m$ out of $r$ modes to be designated as ``fast varying''
ones is inconsequential).

As to the different orientations of the telescope, they affect only sources
external to the telescope, $\mathbf{I}^{I}$, which exclude speckle
modes, $G^{V}$, and dark current, $\mathbf{I}^{D}$. We assume that
there is a known transformation, $R(k):\mathbb{R}^{n}\rightarrow\mathbb{R}^{n}$,
which describes the effect of the telescope roll at frame $k$ on
the otherwise constant image $\mathbf{I}^{I}(0)$, i.e.
\begin{equation}
\mathbf{I}^{I}(k)=R(k)\left\{ \mathbf{I}^{I}(0)\right\} .
\end{equation}
For numerical purposes, we also assume that $R$ is invertible and
differentiable. 

Putting it all together, the problem consists of finding the estimates
for the speckle modes, $\hat{G}^{V}$, the history of their coefficients,
$\left\{ \mathbf{\hat{v}}(k)\right\} $, the history of the jitter
coefficients, $\mathbf{\hat{w}}(k)=\left[\hat{w}_{i,j}(k)\right]_{1\le j\le i\le m}$
and the signal, $\hat{\mathbf{I}}^{I}=\hat{\mathbf{I}}^{I}(0)$. The
latter can be expressed in terms of the former,
\begin{equation}
\hat{\mathbf{I}}^{I}=\mathrm{ramp}\left(\frac{1}{T}\underset{k=1}{\overset{T}{\sum}}R^{-1}(k)\left\{\mathbf{y}(k)-\hat{\mathbf{I}}^{C}(k)-\hat{\mathbf{I}}^{J}(k)-\mathbf{I}^{D}\right\}\right),\label{eq:I_hat_ramp}
\end{equation}
where ``$\mathrm{ramp}$'' is the elementwise ramp function, $\mathbf{y}(k)$
are the intensity measurements and $\hat{\mathbf{I}}^{S}(k)$ and
$\hat{\mathbf{I}}^{J}(k)$ are the estimates of the speckle and jitter
intensities (see eqs.~(\ref{eq:I_avg_def})-(\ref{eq:I_J_def}),(\ref{eq:jitter_modes})),
\begin{alignat*}{1}
\hat{\mathbf{I}}^{S}(k) & =B\cdot\left(\hat{G}^{V}\hat{\mathbf{v}}(k)+G^{U}\mathbf{u}(k)\right)^{\circ2},\\
\hat{\mathbf{I}}^{J}(k) & =B\cdot\underset{i=1}{\overset{m}{\sum}}\left(\underset{j=1}{\overset{i}{\sum}}\hat{w}_{i,j}(k)\hat{\mathbf{g}}_{j}^{V}\right)^{\circ2}.
\end{alignat*}
The optimization problem is then stated as maximzing the log-likelihood
of observing $\left\{ \mathbf{y}(k)\right\} $ assuming a Poisson
distribution, 
\begin{alignat}{1}
\log p\left(\left.\left\{ \mathbf{y}(k)\right\} \right|\hat{G}^{V},\left\{ \mathbf{\hat{v}}(k)\right\} ,\left\{ \mathbf{\hat{w}}(k)\right\} \right) & =\underset{p,k}{\sum}\left(y_{p}(k)\log\hat{I}_{p}(k)-\hat{I}_{p}(k)-\log\left(y_{p}(k)!\right)\right),\label{eq:cost}\\
\hat{\mathbf{I}}(k) & =\hat{\mathbf{I}}^{S}(k)+\hat{\mathbf{I}}^{J}(k)+R(k)\hat{\mathbf{I}}^{I}+\mathbf{I}^{D}.
\end{alignat}

Eliminating $\hat{\mathbf{I}}^{I}$ from the numerical procedure via
Eq.~(\ref{eq:I_hat_ramp}), allows optimizing the cost function
in Eq.~(\ref{eq:cost}) via gradient descent without taking particular
precautions with regard to the relative scaling between the parameters and their
domain (for details, see  Ref.~\citenum{pogorelyuk2019reduced}). For numerical
stability, the $R(k)$ were chosen to be permutation matrices -- switching
pixel places in a manner that best resembles a rotation.

\subsection{Numerical Results}

\begin{figure}
\begin{center}
\begin{tabular}{c}
\includegraphics[width=14cm]{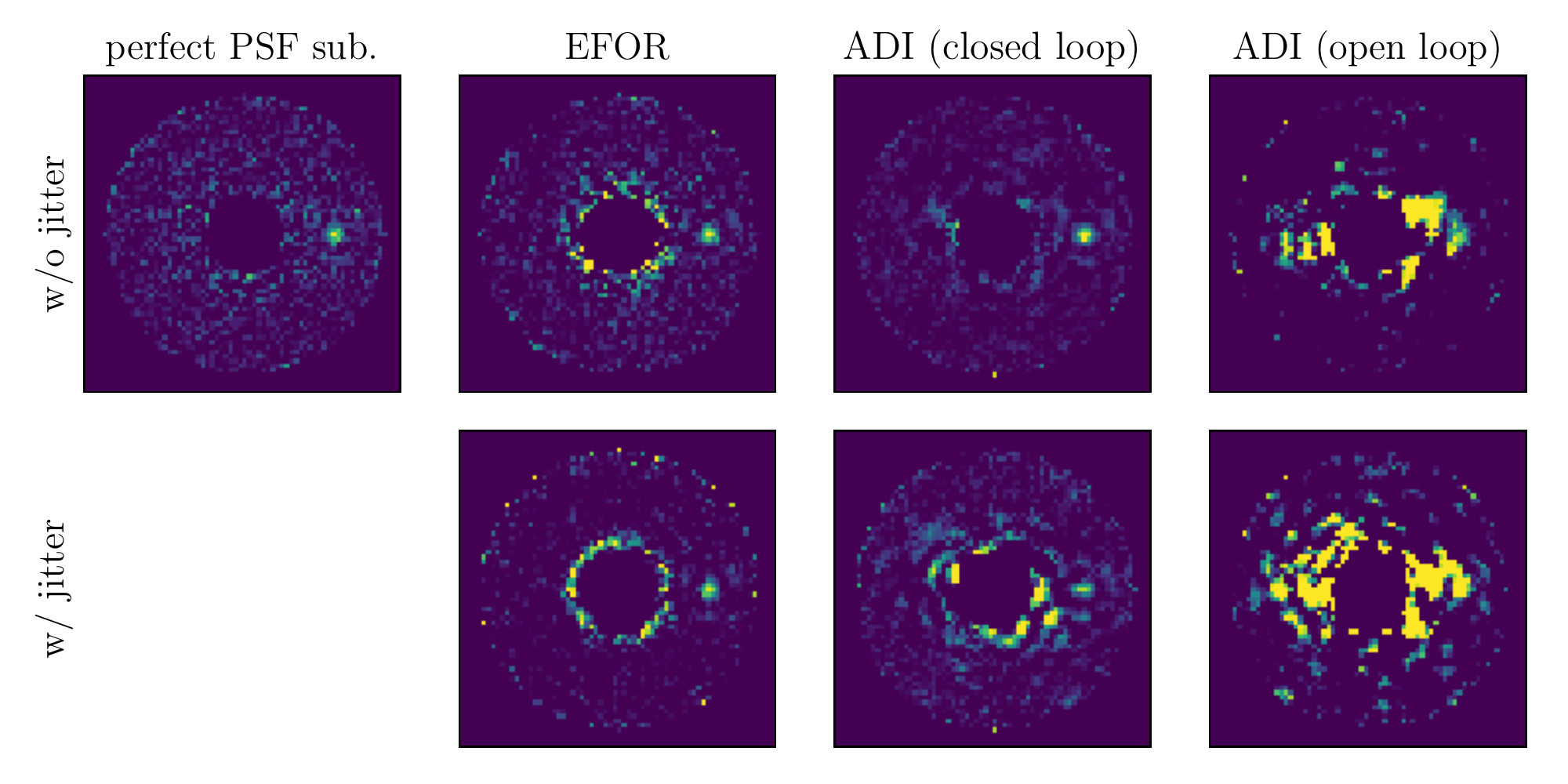}
\end{tabular}
\end{center}
\caption{\label{fig:EFOR_vs_ADI}Incoherent intensity estimates of a planet
at $6\:\lambda/D$, from left to right: Theoretical limit for a $30\:\mathrm{hr}$
long observation with a perfectly known PSF, highest contrast and
no drifts or jitter; Electric Field Order Reduction (EFOR) for a closed-loop $30\:\mathrm{hr}$ long observation with one $26\:\mathrm{deg}$
telescope roll, only speckle (no DM) drift present and jitter either simulated (bottom) or not (top);
Angular Differential Imaging (ADI) for a closed-loop $30\:\mathrm{hr}$
long observation with one $26\:\mathrm{deg}$ telescope roll; ADI
for three open-loop $8\:\mathrm{hr}$ long observation, each starting from a ``perfect'' dark hole and including four $26\:\mathrm{deg}$ telescope rolls.}
\end{figure}

The authors have previously suggested that combining dark hole maintenance with EFOR could lead to smaller errors in post-processing\cite{pogorelyuk2019reduced}. Below, we re-assess this claim in the presence of jitter and high order DM
drift (i.e., when each actuator drifts independently of all others). To this end, we use the data simulated in Sec.~\ref{sub:EKF_numerical} while additionally introducing one $26\:\mathrm{deg}$ telescope roll. A planet with intensity of $5.0\:\frac{\mathrm{\mathrm{photon}}}{\mathrm{frame}}$
(across all wavelengths) was simulated at $6\:\lambda/D$.

Figure~\ref{fig:EFOR_vs_ADI} (left) shows the baseline for post-processing evaluation
-- a simple PSF subtraction assuming that the speckle pattern remains
fixed at its lowest average intensity and is perfectly known. In this ``perfect''
case, the PSF subtraction error is due to shot-noise alone and is
therefore at its theoretical limit. In an open-loop scenario without
DM drift and jitter, the Angular Differential Imaging (ADI) technique
eliminated some, but not all of the speckle (Fig.~\ref{fig:EFOR_vs_ADI}
- right). In the closed-loop scenario ADI and EFOR preformed similarly,
although the residual error of ADI is smoother (Fig.~\ref{fig:EFOR_vs_ADI}
middle).

To quantify the comparison we use the relative post-processing-error
(PPE), based on the error of the estimated incoherent intensity in
the region where the planet's intensity is above its half-max,
\begin{equation}
\mbox{rel. PPE}=\frac{\mbox{avg. post-processing error }}{\mbox{avg. error of perfect PSF subtraction }}.\label{eq:PPF}
\end{equation}
The relative PPE is always greater or equal to $1$ and, if the intensity
of the speckles is uniformly distributed in time, is independent of the duration
of the observation (in the limit of long observations). For the open-loop
scenario we simulated three $8\:\mathrm{hr}$ fixed DM exposures
(four rolls each) with dark hole ``re-creation'' between them
($80\%$ duty cycle), while the closed-loop simulations lasted $30$
hours and the contrast was maintained throughout. The potentially
adverse effects of speckle drift induced by pointing maneuvers were
not simulated.

Closing the loop always led to better relative PPEs, as illustrated in table~\ref{tab:results} which is based on multiple realizations of each combination of a post-processing
method and an observation scenario. Lowering the shot noise alone (see Sec.~\ref{sub:EKF_numerical}), reduces the errors by at least a factor of $2$. At the intermediate angular separation of $6\:\lambda/D$,
ADI and EFOR performed similarly well, with the latter being able to
identify and subtract the jitter modes when present. For lower angular separations, the PPE depends
strongly on the jitter profile and roll sequences, although EFOR doesn't
suffer accuracy losses as much as ADI (this suggests that WFIRST has large sensitivity to jitter, although its quantitative analysis is beyond the scope of this paper). EFOR, however, breaks down in
the presence of drift of statistically independent DM actuators since the resulting perturbations
of the electric field are not low-order and have a larger amplitude
than those of the speckles drift (arguably, this is only the case because DM drift is large compared to the low order Zernike drift).

\begin{table}[ht]
\begin{center}
\begin{tabular}{|c|c|c|c|}
\hline
\rule[-1ex]{0pt}{3.5ex}
 & monochromatic & monochromatic & broadband\tabularnewline
 & w/o DM drift & w/o DM drift & w/ DM drift\tabularnewline
 & w/o jitter & w/ jitter & w/ jitter\tabularnewline
\hline
\rule[-1ex]{0pt}{3.5ex}
open loop + ADI & 3.0 & 4.5 & 12\tabularnewline
\hline
\rule[-1ex]{0pt}{3.5ex}
closed loop +ADI & \textbf{1.5} & 2.0 & \textbf{5}\tabularnewline
\hline
\rule[-1ex]{0pt}{3.5ex}
closed loop +EFOR & \textbf{1.5} & \textbf{1.5} & 15\tabularnewline
\hline 
\end{tabular}
\end{center}
\caption{\label{tab:results}Post-processing errors relative to PSF subtraction
under perfect conditions (Eq.~(\ref{eq:PPF})) for various observation
strategies (fixed DMs with periodic re-creation of the dark hole vs.
closed-loop), instability sources (only drift in the first 21 Zernike modes, additional
slowly varying residual jitter, additional high order DM actuators drift), detector
types (IFS vs. broadband detectors) and post-processing algorithms
(ADI vs. EFOR). Closed-loop strategy is always preferable to open-loop
and ADI generally gets close to the theoretical detection limit. EFOR
has a slight advantage since it is capable of identifying and subtracting
``jitter modes'' as long as the drift is not ``high-order'' as
is the case in the pretense of DM drift.}
\end{table}

\section{Conclusions}

In this work we reaffirmed the benefits of actively maintaining high contrast throughout coronagraphic observations in the presence of jitter, DM actuators drift and broadband detectors. Numerical simulations of WFIRST
suggest that regardless of the observation parameters, closing the loop on the electric field in the dark hole reduces shot noise and therefore significantly increases the post-processing accuracy.

A continuous closed-loop observation strategy would allow detecting fainter planets compared to periodic re-creation of the dark hole by pointing the telescope at bright reference stars. Although not included in our analysis, slewing the telescope back and forth between target and reference stars might introduce thermal stresses which would negatively impact the stability of the wavefront. Active dark hole maintenance avoids such maneuvers and the lower duty cycle and destabilization risks associated with them.

In Sec.~\ref{sec:maintenance} we've shown that the averaged pointing
jitter manifests itself as an incoherent source, while DM actuators
drift is similar to high-order instabilities in other optical components
(e.g., the primary mirror). Both effects decrease the contrast in the
dark hole, but are implicitly tackled by existing batch and recursive
electric field estimation algorithms. These approaches can be slightly
modified to process broadband measurements by expressing them as an
incoherent sum over several narrowband intensities.

Assuming that the jitter is well modelled by fast perturbations in
just a few electric field modes (e.g., tip and tilt), they can be
estimated together with the low-order speckle drift modes in post-processing
via EFOR (accounting for possible telescope rolls, see Sec.~\ref{sec:EFOR}).
The jitter effects can then be subtracted in post-processing, resulting
in a slightly better performance compared to ADI. However,  EFOR is not applicable when the wavefront errors are dominated by high-order disturbances due to DM drift. In that case, the lowest post-processing error was achieved by a combination of dark hole maintenance and ADI.

\acknowledgments 
This work was partially supported by the Army Research Office, award number W911NF-17-1-0512.

\appendix
\section{\label{sec:appendix}Broadband Pair-probing and EFC Conditioning}

Below, we analyze the numerical conditioning of the pair-probing broadband
electric field estimate in Eq.~(\ref{eq:probing_estimate}) and the
corresponding EFC control, Eq.~(\ref{eq:probing_control}). Even
though the latter is based on the former, there is a large discrepancy
between their ``relative errors'' as depicted in Fig.~\ref{fig:EFC}(b).
To see why this is the case, it is sufficient to consider a dark
hole with a single ($n=1$) broadband detector ($m=1$ and $l>1$).

With $n=m=1$, eqs.~(\ref{eq:probing_I}) and (\ref{eq:probing_estimate})
become 
\begin{alignat}{1}
I^{DH}(\pm\delta\mathbf{u}_{j}) & =\left(\mathbf{E}^{DH}\pm G^{U}\delta\mathbf{u}_{j}\right)^{T}\left(\mathbf{E}^{DH}\pm G^{U}\delta\mathbf{u}_{j}\right)+I^{DIJ}\in\mathbb{R},\\
4\begin{bmatrix}\left(G^{U}\delta\mathbf{u}_{1}\right)^{T}\\
\vdots\\
\left(G^{U}\delta\mathbf{u}_{s}\right)^{T}
\end{bmatrix}\hat{\mathbf{E}}^{DH} & \approx\begin{bmatrix}I^{DH}(+\delta\mathbf{u}_{1})-I^{DH}(-\delta\mathbf{u}_{1})\\
\vdots\\
I^{DH}(+\delta\mathbf{u}_{s})-I^{DH}(-\delta\mathbf{u}_{s})
\end{bmatrix}\in\mathbb{R}^{s},\label{eq:probing_estimate_single_pixel}
\end{alignat}
respectively. Denoting, $\delta C=\begin{bmatrix}\delta\mathbf{u}_{1} & \cdots & \delta\mathbf{u}_{s}\end{bmatrix}\in\mathbb{R}^{a\times s}$
with $a$ standing for the number of DM actuators, (\ref{eq:probing_estimate_single_pixel})
can be rewritten as
\begin{equation}
4\left(G^{U}\delta C\right)^{T}\hat{\mathbf{E}}^{DH}\approx\mathbf{b},\label{eq:least_squares_probing}
\end{equation}
with $G^{U}\in\mathbb{R}^{2l\times a}$, $\hat{\mathbf{E}}^{DH}\in\mathbb{R}^{2l}$ and $\mathbf{b}$ defined as the right-hand side of (\ref{eq:probing_estimate_single_pixel}).
The accuracy of $\hat{\mathbf{E}}^{DH}$ therefore depends on the
condition number of $G^{U}\delta C$ which may be very large due to
the ambiguity between speckles at different wavelengths. In particular,
the rows of $G^{U}$which correspond to ``nearby'' wavelengths are
``almost'' collinear, a fact which was not accounted for in
Sec.~\ref{sub:dark_hole_creation}. 

Let
\begin{flalign}
G^{U} & =U_{G}\Sigma_{G}V_{G}^{T},\label{eq:G_U_svd}\\
\delta C & =U_{C}\Sigma_{C}V_{C}^{T}
\end{flalign}
be the singular value decompositions of $G^{U}$ and $\delta C$ with
$U_{G},\Sigma_{G}\in\mathbb{R}^{2l\times 2l}$, $V_{G}\in\mathbb{R}^{a\times 2l}$,
$U_{C}\in\mathbb{R}^{a\times s}$ and $\Sigma_{C},V_{C}\in\mathbb{R}^{s\times s}$.
The least squares solution of (\ref{eq:least_squares_probing}) is
given by,
\begin{equation}
\hat{\mathbf{E}}^{DH}=\frac{1}{4}U_{G}\Sigma_{G}^{-1}\left(U_{C}^{T}V_{G}\right)^{+}\Sigma_{C}^{-1}V_{C}^{T}\mathbf{b}\label{eq:E_DH_estimate}
\end{equation}
where $\left(U_{C}^{T}V_{G}\right)^{+}$ is the pseudo-inverse of $U_{C}^{T}V_{G}$.
Due to the above mentioned spectral ambiguity, the singular values
of $G^{U}$ (on the main diagonal of $\Sigma_{G}$) may vary by several
orders of magnitude. Hence, relatively small errors in $\mathbf{b}$, when
amplified by $\Sigma_{G}^{-1}$, yield large errors in the estimate,
$\hat{\mathbf{E}}^{DH}$, as illustrated in Fig.~\ref{fig:EFC}(b).

However, to compute the control, $\Delta\mathbf{u}^{DH}$, the estimate
is multiplied by the EFC gain in Eq.~(\ref{eq:probing_control}).
This gain is given by
\begin{equation}
K_{EFC}=\left(\left(G^{U}\right)^{T}G^{U}+\mu I\right)^{-1}\left(G^{U}\right)^{T},
\end{equation}
where $\mu>0$ is a regularization constant which makes $\left(G^{U}\right)^{T}G^{U}+\mu I$
well conditioned. Together with eqs.~(\ref{eq:G_U_svd}) and (\ref{eq:E_DH_estimate}), the
control becomes
\begin{equation}
\Delta\mathbf{u}^{DH}=-K_{EFC}\hat{\mathbf{E}}^{DH}=-\frac{1}{4}\left(V_{G}\Sigma_{G}^{2}V_{G}^{T}+\mu I\right)^{-1}V_{G}\left(U_{C}^{T}V_{G}\right)^{+}\Sigma_{C}^{-1}V_{C}^{T}\mathbf{b},
\end{equation}
and does not contain the ill-conditioned $\Sigma_{G}^{-1}$. Indeed,
$\Delta\mathbf{u}^{DH}$ is not severely affected by measurement
errors and non-linearities: in the numerical results of Sec.~\ref{sub:dark_hole_creation},
each additional pixel increases the relative error in $\Delta\mathbf{u}^{DH}$
by about $3\%$ (and since there are $n=2608$ pixels, the total relative
error is about $\sqrt{2608}\cdot 0.03\approx1.5$, as seen in Fig.~\ref{fig:EFC}(b)).


\bibliographystyle{spiejour}   

\begin{thebibliography}{10}

\bibitem{stark2019exoearth}
C.~C. Stark {\em et~al.}, ``Exoearth yield landscape for future direct imaging
  space telescopes,'' {\em Journal of Astronomical Telescopes, Instruments, and
  Systems} {\bf 5}(2), 1 -- 20  (2019).

\bibitem{seo2019testbed}
B.-J. Seo {\em et~al.}, ``Testbed demonstration of high-contrast coronagraph
  imaging in search for earth-like exoplanets,'' in {\em Techniques and
  Instrumentation for Detection of Exoplanets IX},  S.~B. Shaklan, Ed., {\em
  Proc.SPIE} {\bf 11117}, 599 -- 609, International Society for Optics and
  Photonics, SPIE  (2019).

\bibitem{douglas2018wfirst}
E.~S. Douglas {\em et~al.}, ``Wfirst coronagraph technology requirements:
  status update and systems engineering approach,'' in {\em Modeling, Systems
  Engineering, and Project Management for Astronomy {VIII}},   {\bf 10705},
  1070526, International Society for Optics and Photonics  (2018).

\bibitem{demers2018review}
R.~Demers, ``Review and update of {WFIRST} coronagraph instrument design and
  technology (conference presentation),'' in {\em Space Telescopes and
  Instrumentation 2018: Optical, Infrared, and Millimeter Wave},  {\em
  Proc.SPIE} {\bf 10698}  (2018).

\bibitem{nemati2017sensitivity}
B.~Nemati, J.~E. Krist, and B.~Mennesson, ``Sensitivity of the {WFIRST}
  coronagraph performance to key instrument parameters,'' in {\em Techniques
  and Instrumentation for Detection of Exoplanets {VIII}},  {\em Proc.SPIE}
  {\bf 10400}, 1040007, International Society for Optics and Photonics  (2017).

\bibitem{shaklan2011stability}
S.~B. Shaklan {\em et~al.}, ``Stability error budget for an aggressive
  coronagraph on a 3.8 m telescope,'' in {\em Techniques and Instrumentation
  for Detection of Exoplanets {V}},  {\em Proc.SPIE} {\bf 8151}, 815109,
  International Society for Optics and Photonics  (2011).

\bibitem{zhou2018high}
H.~Zhou {\em et~al.}, ``High accuracy coronagraph flight {WFC} model for
  {WFIRST-CGI} raw contrast sensitivity analysis,'' in {\em Space Telescopes
  and Instrumentation 2018: Optical, Infrared, and Millimeter Wave},
  M.~Lystrup, H.~A. MacEwen, G.~G. Fazio, {\em et~al.}, Eds.,  {\bf 10698}, 811
  -- 824, International Society for Optics and Photonics, SPIE  (2018).

\bibitem{garreth2017performance}
G.~Ruane {\em et~al.}, ``Performance and sensitivity of vortex coronagraphs on
  segmented space telescopes,'' in {\em Techniques and Instrumentation for
  Detection of Exoplanets VIII},  S.~Shaklan, Ed., {\em Proc.SPIE} {\bf 10400},
  140 -- 155, International Society for Optics and Photonics, SPIE  (2017).

\bibitem{mennesson2016habitable}
B.~Mennesson {\em et~al.}, ``The habitable exoplanet ({HabEx}) imaging mission:
  preliminary science drivers and technical requirements,'' in {\em Space
  Telescopes and Instrumentation 2016: Optical, Infrared, and Millimeter Wave},
   H.~A. MacEwen, G.~G. Fazio, M.~Lystrup, {\em et~al.}, Eds., {\em Proc.SPIE}
  {\bf 9904}, 212 -- 221, International Society for Optics and Photonics, SPIE
  (2016).

\bibitem{bolcar2017large}
M.~R. Bolcar {\em et~al.}, ``The large uv/optical/infrared surveyor ({LUVOIR}):
  Decadal mission concept design update,'' in {\em UV/Optical/IR Space
  Telescopes and Instruments: Innovative Technologies and Concepts VIII},
  H.~A. MacEwen and J.~B. Breckinridge, Eds., {\em Proc.SPIE} {\bf 10398}, 79
  -- 102, International Society for Optics and Photonics, SPIE  (2017).

\bibitem{bailey2018lessons}
V.~P. Bailey {\em et~al.}, ``Lessons for wfirst cgi from ground-based
  high-contrast systems,'' in {\em Space Telescopes and Instrumentation 2018:
  Optical, Infrared, and Millimeter Wave},   {\bf 10698}, 106986P,
  International Society for Optics and Photonics  (2018).

\bibitem{give2011pair}
A.~Give'on, B.~D. Kern, and S.~B. Shaklan, ``Pair-wise, deformable mirror,
  image plane-based diversity electric field estimation for high contrast
  coronagraphy,'' in {\em Techniques and Instrumentation for Detection of
  Exoplanets {V}},  {\em Proc.SPIE} {\bf 8151}, 815110, International Society
  for Optics and Photonics  (2011).

\bibitem{give2007electric}
A.~Give'on {\em et~al.}, ``Electric field conjugation-a broadband wavefront
  correction algorithm for high-contrast imaging systems,'' in {\em Bulletin of
  the American Astronomical Society},   {\bf 39}, 975  (2007).

\bibitem{krist2018wfirst}
J.~Krist {\em et~al.}, ``{WFIRST coronagraph flight performance modeling},'' in
  {\em Space Telescopes and Instrumentation 2018: Optical, Infrared, and
  Millimeter Wave},  M.~Lystrup, H.~A. MacEwen, G.~G. Fazio, {\em et~al.},
  Eds., {\em Proc.SPIE} {\bf 10698}, 788 -- 810, International Society for
  Optics and Photonics, SPIE  (2018).

\bibitem{pogorelyuk2019dark}
L.~Pogorelyuk and N.~J. Kasdin, ``Dark hole maintenance and a posteriori
  intensity estimation in the presence of speckle drift in a high-contrast
  space coronagraph,'' {\em The Astrophysical Journal} {\bf 873}, 95  (2019).

\bibitem{shi2019wfirst}
F.~Shi {\em et~al.}, ``{WFIRST low order wavefront sensing and control
  ({LOWFS}/{C}) performance on line-of-sight disturbances from multiple
  reaction wheels},'' in {\em Techniques and Instrumentation for Detection of
  Exoplanets IX},  S.~B. Shaklan, Ed., {\em Proc.SPIE} {\bf 11117}, 170 -- 178,
  International Society for Optics and Photonics, SPIE  (2019).

\bibitem{prada2019high}
C.~M. Prada, E.~Serabyn, and F.~Shi, ``{High-contrast imaging stability using
  {MEMS} deformable mirror},'' in {\em Techniques and Instrumentation for
  Detection of Exoplanets IX},  S.~B. Shaklan, Ed., {\em Proc.SPIE} {\bf
  11117}, 112 -- 118, International Society for Optics and Photonics, SPIE
  (2019).

\bibitem{pogorelyuk2019reduced}
L.~Pogorelyuk, N.~J. Kasdin, and C.~W. Rowley, ``Reduced order estimation of
  the speckle electric field history for space-based coronagraphs,'' {\em The
  Astrophysical Journal} {\bf 881}, 126  (2019).

\bibitem{lowrance1998coronagraphic}
P.~J. Lowrance {\em et~al.}, ``A coronagraphic search for substellar companions
  to young stars,'' in {\em NICMOS and the VLT: A New Era of High Resolution
  Near Infrared Imaging and Spectroscopy},  W.~Freudling and R.~Hook, Eds.,
  {\em European Southern Observatory Conference and Workshop Proceedings} {\bf
  55}, 96, European Southern Observatory  (1998).

\bibitem{marois2006angular}
C.~Marois {\em et~al.}, ``Angular differential imaging: A powerful
  high-contrast imaging technique,'' {\em The Astrophysical Journal} {\bf 641},
  556--564  (2006).

\bibitem{jovanovic2018review}
N.~Jovanovic {\em et~al.}, ``Review of high-contrast imaging systems for
  current and future ground-based and space-based telescopes: Part {II}. common
  path wavefront sensing/control and coherent differential imaging,'' in {\em
  Adaptive Optics Systems {VI}},  {\em Proc.SPIE} {\bf 10703}  (2018).

\bibitem{riggs2014optimal}
A.~E. Riggs, N.~J. Kasdin, and T.~D. Groff, ``Optimal wavefront estimation of
  incoherent sources,'' in {\em Space Telescopes and Instrumentation 2014:
  Optical, Infrared, and Millimeter Wave},  {\em Proc.SPIE} {\bf 9143}, 914324,
  International Society for Optics and Photonics  (2014).

\bibitem{pogorelyuk2020efficient}
L.~Pogorelyuk, C.~W. Rowley, and N.~J. Kasdin, ``An efficient approximation of
  the kalman filter for multiple systems coupled via low-dimensional stochastic
  input,'' {\em arXiv preprint arXiv:1911.10443}   (2019).

\bibitem{riggs2018fast}
A.~E. Riggs {\em et~al.}, ``Fast linearized coronagraph optimizer ({FALCO})
  {I}: a software toolbox for rapid coronagraphic design and wavefront
  correction,'' in {\em Space Telescopes and Instrumentation 2018: Optical,
  Infrared, and Millimeter Wave},  H.~A. MacEwen, M.~Lystrup, G.~G. Fazio, {\em
  et~al.}, Eds., {\em Proc.SPIE} {\bf 10698}, {SPIE}  (2018).

\bibitem{mawet2011dim}
D.~Mawet, B.~Mennesson, E.~Serabyn, {\em et~al.}, ``A dim candidate companion
  to {Epsilon Cephei},'' {\em The Astrophysical Journal Letters} {\bf 738}(1),
  L12  (2011).

\bibitem{soummer2012detection}
R.~Soummer, L.~Pueyo, and J.~Larkin, ``Detection and characterization of
  exoplanets and disks using projections on {Karhunen-Lo\`eve} eigenimages,''
  {\em The Astrophysical Journal Letters} {\bf 755}, L28  (2012).

\bibitem{amara2012pynpoint}
A.~Amara and S.~P. Quanz, ``{PYNPOINT}: An image processing package for finding
  exoplanets,'' {\em Monthly Notices of the Royal Astronomical Society} {\bf
  427}, 948--955  (2012).

\bibitem{ren2018non}
B.~Ren {\em et~al.}, ``Non-negative matrix factorization: Robust extraction of
  extended structures,'' {\em The Astrophysical Journal} {\bf 852}, 104
  (2018).

\bibitem{pogorelyuk2019maintaining}
L.~Pogorelyuk and N.~J. Kasdin, ``Maintaining a dark hole in a high contrast
  coronagraph and the effects of speckles drift on contrast and post processing
  factor,'' in {\em Techniques and Instrumentation for Detection of Exoplanets
  IX},  S.~B. Shaklan, Ed., {\em Proc.SPIE} {\bf 11117}, 397 -- 403,
  International Society for Optics and Photonics, SPIE  (2019).

\end{thebibliography}

\listoffigures
\listoftables

\end{spacing}
\end{document}